\documentstyle[12pt]{article}
\begin{document}

\vspace*{0.5cm}
\baselineskip=0.8cm
\begin{center}
{\Large\bf Spin-Density-Functional Theory of Circular and Elliptical 
Quantum Dots}
\vspace{2.0cm}

{\large\bf Kenji Hirose$^{\dagger}$ and Ned S. Wingreen}
\vspace{0.7cm}

NEC Research Institute, 4 Independence Way, Princeton, New Jersey
08540\\
\vspace{0.5cm}
\end{center}

\baselineskip=0.69cm

\vspace{1.5cm}

\begin{center}
{\large\bf Abstract}
\end{center}

\vspace{0.6cm}

Using spin-density-functional theory, we study the electronic states of
a two-dimensional parabolic quantum dot with up to $N=58$ electrons.
We observe a shell structure for the filling of the dot with electrons.
Hund's rule determines the spin configuration of the ground state, but
only up to 22 electrons. At specific $N$, the ground state is degenerate, 
and a small elliptical deformation of the external potential induces a 
rotational charge-density-wave (CDW) state.  Previously identified 
spin-density-wave (SDW) states are shown to be artifacts of broken 
spin symmetry in density-functional theory.

\vspace{1.0cm}

\ \\


\newpage

Quantum dots have recently attracted much interest both experimentally
and theoretically. One realization of a quantum dot is a small island 
fabricated in a two-dimensional electron gas laterally confined by an
external
potential and containing a few to a few hundred electrons\cite{Kastner}.
Experimentally, measuring the tunnel conductance\cite{Meirav} and
capacitance
\cite{Ashoori} by changing the gate voltage attached to the quantum dot, 
one observes a peak every time 
the average number of electrons 
increases by one. The spacing of peaks, or addition spectrum, reflects
the energy 
differences between ground states of the dot with different numbers
of electrons. Each disordered dot has its own characteristic addition
spectrum, but recently it has become possible to fabricate dots
so clean that the addition spectra are reproducible from dot to
dot\cite{Tarucha}. 
Among the features of these clean, parabolic dots 
are atomic-like shell structures, Hund's rules, and reproducible 
transition rates\cite{Tarucha}.

The advent of atomic-like spectra in quantum dots calls for 
appropriately quantitative theoretical tools. Presently,  
exact diagonalization of the full Hamiltonian is limited 
to a small number of electrons in the dot\cite{Maksym,Eto97}.
Thomas-Fermi, Hartree\cite{Kumar}, and Hartree-Fock methods 
\cite{Fujito,Koonin} all suffer from sizeable systematic errors.
Here, we treat the electronic states of a dot using the
spin-density-functional method\cite{Hess,Stopa,Koskinen}.
We find shell structures in the
addition-energy spectrum for a circular, parabolic external potential.
Hund's rule determines the ground-state spin configurations, but only up
to 22
electrons.  Elliptically deforming the external potential 
eliminates the shell structures, and Hund's rule is replaced
by a more Pauli-like behavior of the total spin.
At specific N, the ground state is degenerate,
 and a small elliptical deformation of the external potential
induces a rotational charge-density-wave (CDW) state.
The spin-density-wave (SDW) states found by Koskinen, Manninen, and 
Riemann (KMR)\cite{Koskinen} are artifacts of broken spin symmetry 
in density-functional theory.

We solve the following Kohn-Sham equations numerically for
a two-dimensional parabolic quantum dot,
and iterate until self-consistent solutions are obtained\cite{Kohn};
\begin{eqnarray}
&&\left[-\frac{\hbar^2}{2m^*}\nabla^2
+\frac{e^2}{\kappa}\int\frac{\rho({\bf r'})}{|{\bf r-r'}|}d{\bf
r'}+\frac{\delta E_{\rm xc}[\rho,\zeta]}{\delta\rho^{\sigma}({\bf r})}
+\frac{1}{2}m^*\omega_0^2r^2\right]\Psi_i^{\sigma}({\bf r})
=\epsilon_i^{\sigma}\Psi_i^{\sigma}({\bf r})\\
&&\hspace*{2.0cm}\rho({\bf r})=\sum_{\sigma}\rho^{\sigma}({\bf r})
=\sum_{\sigma}\sum_{i}|\Psi_i^{\sigma}({\bf r})|^2.
\end{eqnarray}
Here $\sigma$ denotes the spin index, $\zeta({\bf r})$ is the local
spin polarization, and $E_{\rm xc}$ is the exchange-correlation energy
functional, for which we use the local-density
approximation\cite{Tanatar}
\begin{eqnarray}
&&E_{\rm xc}=\int\rho({\bf r})\epsilon_{\rm xc}
[\rho({\bf r}),\zeta({\bf r})] d{\bf r}\\
&&\zeta({\bf r})=\frac{\rho^{\uparrow}({\bf r})-\rho^{\downarrow}
({\bf r})}{\rho({\bf r})}.
\end{eqnarray}
To solve the equation, we expand the 
$\Psi_i^{\sigma}({\bf r})$ in a Fock-Darwin representation;\\
$\displaystyle{\varphi_{n,k}^{\sigma}(r,\theta)=
|n,k\rangle =\sqrt{\frac{n!}{2\pi(n+|k|)!}}
\frac{1}{l}e^{-\frac{r^2}{4l^2}}(\frac{r^2}{2l^2})^{\frac{|k|}
{2}}{\rm L}^{|k|}_n(\frac{r^2}{2l^2})e^{-ik\theta}\chi_{\sigma}}$,
where ${\rm L}^{|k|}_{n}(x)$ is a Laguerre polynomial,
$\displaystyle{l=\sqrt{\hbar/
(2m^{*}\omega_0)}}$ and $\chi_{\sigma}$ is a spin function.
The non-interacting, single-particle levels form a ladder, 
$\epsilon_{n,k}=(2n+|k|+1)\hbar\omega_0 = M \hbar \omega$, 
with rung degeneracy $M$.  The ground-state energy of a quantum 
dot with N electrons is obtained from \begin{equation}
E(N)=\sum_{i,\sigma}\epsilon_i^{\sigma}-\frac{e^2}{2\kappa}\int\frac{\rho({\bf
r})\rho({\bf r'})}{|{\bf r-r'}|}d{\bf r}d{\bf r'}-\sum_{\sigma}\int
\rho^{\sigma}({\bf r})\frac{\delta E_{\rm xc}[\rho,\zeta]}
{\delta\rho^{\sigma}({\bf r})}d{\bf r}+E_{\rm xc}.
\end{equation}
We use the material constants for GaAs, $m^*=0.067m, \kappa=12.9$, and
the external potential is fixed at $\hbar \omega_0=3.0$ meV.
The resulting dimensionless interaction strength is
$(e^2/\kappa\ell_0)/\hbar \omega_0 = 1.9$, where 
$\ell_0 = \sqrt{\hbar/(m^*\omega_0)}$. 

{\it Shell structure} -- At low temperatures, electron hopping into a dot
containing $N$ electrons is suppressed except when the 
ground-state energy $E(N)$ is equal to $E(N+1)$.
This degeneracy condition determines the observed conductance
oscillation peaks which occur at the chemical potentials
$\mu(N+1)=E(N+1)-E(N)$.
The addition energy $\Delta(N)$ needed to put an extra electron
in the dot is obtained from
$\Delta(N)=\mu(N+1)-\mu(N)=E(N+1)-2E(N)+E(N-1)$.
Fig. 1(a) shows the addition energy $\Delta(N)$ as a function of
electron
number $N$ for a circular, parabolic potential.
The dotted line indicates $\Delta(N)$ obtained from a
classical electrostatic analysis with no kinetic energy\cite{Shikin}.
Overall, $\Delta(N)$ decreases with $N$, as the dot and its
capacitance grow. On average, the addition energy obtained from
the density-functional calculation is close to the
classical 
electrostatic result $e^2/C$.
However, we see small zig-zag structures, and large peaks 
at electron numbers $N=2,6,12,20,30,42$, and $56$.
In the single-particle spectrum for a parabolic potential, 
the electronic states of the dot form
closed shell structures at these numbers.
Even in the presence of electron-electron interaction, 
extra energy is required to add
one more electron to a closed shell.
The peak heights decrease as the number $N$ increases,
consistent with recent experiments which observed
peaks in the addition energy up to 12 electrons\cite{Tarucha}.  

{\it Hund's rule} -- 
By analogy to atoms, we expect that Hund's rule for total spin
will apply in the present situation. According 
to Hund's rule, as degenerate
states are filled, the total spin $S$ takes the maximum value allowed
by the exclusion principle and becomes zero for closed shells.
Fig. 2(a) shows the spin configuration as a function of the electron
number $N$ for the circular, parabolic potential.
The dotted line represents the spin configuration when Hund's rule is
satisfied. We can see that, for up to 22 electrons filling the dot, 
the spin configurations obey Hund's rule\cite{Tarucha}.
For larger dots, Hund's rule is
violated and the high spin states are suppressed. 
In particularly, the total spin becomes zero at
electron numbers $N=24,34,46$ instead of the expected $S=2$.
In the upper inset of Fig.1, we show the total energy $E(N)$
as a function of the total spin for these states.
At $N=16$, the $S=2$ state is 0.10 meV lower in energy than the 
$S=0$ state, which follows Hund's rule.
In contrast, at $N=24$ the $S=0$ state is 0.05 meV lower in energy
than the $S=2$ state. This trend is enhanced as the number increases.
These energy differences are sufficiently small that weak magnetic
fields, of order 300 gauss, will favor an  $S=2$ ground state.

The breakdown of Hund's rule is due to the non-parabolic 
effective potential caused by Coulomb interactions. 
For example, without interactions $N=24$ corresponds to 
20 electrons in filled inner shells and four 
``valence" electrons distributed among 10 degenerate 
states: $|n,k\rangle$ $=$ $|0,\pm 4\rangle$, $|1,\pm 2\rangle$, and
$|2,0\rangle$,
spin up and down. Coulomb interactions
deform the radial potential and lower the energy of the
single-particle states with larger angular momentum $|k|$. 
The system could minimize its
{\it exchange} energy by creating an $S=2$ state, ${\it i.e.}$, putting all
four valence electrons into spin up states.
Instead, for $N=24$, the system minimizes its {\it single-particle} energy 
by putting all four electrons into $k=\pm 4$ states,
giving a total spin $S=0$, and breaking Hund's rule.

To confirm this result, 
we performed an exact diagonalization of $N=4$
valence electrons in a restricted basis set of eight states: 
$|0,\pm 4\rangle$ and $|1,\pm 2\rangle$, spin up and down. 
The Hamiltonian we employed is
\begin{equation}
H=\sum^{{\rm N=4}}_{i=1}\left[ -\frac{\hbar^2}{2m^*}\nabla_i^2
+\frac{1}{2}m^*\omega_0^2r_i^2+\gamma r_i^4\right]
+\sum_{i<j}\frac{e^2}{\kappa|{\bf r}_i-{\bf r}_j|},
\label{nonpar}
\end{equation}
where the ${\gamma r_i^4}$ term is introduced to split the degenerate
single-particle energies $\epsilon_{0,\pm 4}$ and $\epsilon_{1,\pm 2}$.
The resulting eigenstates of the four electrons can by labelled by total
spin $S$ and $S_z$
and total angular momentum $L_z$.
In the lower inset of Fig. 1, 
we have plotted the total energy as a function of
$\Delta=\epsilon_{1,\pm 2}-\epsilon_{0,\pm 4}$.
If the splitting $\Delta$ is small, the ground state is $S=2, L_z=0$,
consistent with Hund's rule. But for $\Delta$ larger than
$1.4 \,\rm{meV}$, the ground state becomes $S=0, L_z=0$, indicating a
violation of
Hund's rule.

{\it Spin-density-wave states} --
For a dot with circular symmetry, the eigenstates can always
be chosen to have definite angular momentum $L_z$, and hence
circularly symmetric charge density. Nevertheless, KMR\cite{Koskinen} 
reported recently on a 
spontaneous breaking of circular symmetry in a 
spin-density-functional calculation of a parabolic quantum dot. 
Indeed, we confirm that Eqs. (1-5) yield spin-density-wave (SDW) ground states
at particular numbers of electrons, {\it e.g.} $N=24, 34$, as reported in
Ref. \cite{Koskinen}. These are precisely the $S=0$ ground states 
discussed above in the context of breaking of Hund's rule.
Within spin-density-functional theory, even for $S=0$, the system lowers
its exchange energy slightly by mixing in $k=\pm 2$ states with the 
lower lying $k=\pm 4$ orbitals. The result is an SDW state. However,
from our exact diagonalization studies with $N=4$ in the  restricted
basis set, we find that the SDW states are 
due to an unphysical mixture between states of different
total spin: $S=0, S_z=0$ and $S=1, S_z=0$. Hence, the SDW states are
artifacts
of the well known difficulty of spin-density-functional theory that only
the
$S_z$ component of total spin can be specified.  We conclude that the
correct  
ground states for $N=24, 34,$ and $46$ have $S=0, L_z=0$ and retain
circularly symmetry.

{\it Charge-density-wave states} --
We also find that for certain $N$  (cf. Table  1), Eqs. (1-5)
predict a rotational
charge-density wave (CDW) near the edge of the dot. 
Fig. 3 shows an example of such a CDW state for $N=31$. 
The numbers shown in bold face in Table 1 indicate a strong spin-density
modulation, as for $\rho^{\uparrow}({\bf r})$ at $N=31$, 
while a weaker modulation occurs in the opposite spin density. 
The numbers in bold face correspond to a closed shell
plus one electron, indicating that the extra electron is added to 
the lowest orbital in the next shell, namely the one with highest
angular momentum. By circular symmetry, this
orbital is doubly degenerate. Hence the ground state of the
entire dot is doubly degenerate.
(In contrast with atoms, the spin-orbit
interaction in GaAs dots is too small to split this 
degeneracy\cite{Darnhofer}.) The spin-density-functional result
is a mixture of these two degenerate ground states. 
For example, at N=31 the total angular
momentum will be $L_z = \pm 5$, giving a charge density modulation
$\ \ \sim |\exp(i 5 \theta) + \exp(-i 5 \theta)|^2 \sim \cos^2(5\theta)$
as observed
in Fig. 3. 

We have investigated the charge density $\rho^{\sigma}({\bf r})$
for $N=3$ by exact diagonalization to confirm the above interpretation.
As expected, we find that
there are two degenerate ground states, with $L_z=\pm 1$,
and that a coherent mixture of these states produces 
almost exactly the same charge density obtained in the
density-functional
calculation.

{\it Elliptical dots} --
To investigate the effect of removing circular symmetry,
we consider elliptically deformed potentials 
$\displaystyle{V({\bf
r})=m^*(\omega^2_xx^2+\omega^2_yy^2)}/2$, with 
$\displaystyle{\omega_0^2=(\omega_x^2+\omega_y^2)/2}$, which lift
the large degeneracies of the shell structure. 
Fig.1(b) and (c) show that as the deformation grows
the regular zig-zag pattern found for the circular 
potential becomes irregular and the large peaks at large $N$
disappear. One can still see large peaks at $N=2,6,12$, and $20$
electrons in (b), which are the remnants of the closed-shell structures.
However, in (c) such large peaks are present only at $N=2$ and $6$.

Fig. 2(b) and (c) show the spin configurations for the same
deformed external potentials. We can see that Hund's
rule is satisfied up to $N=15$ in (b) but only up to only $N=8$
electrons in (c).
The high spin states are suppressed
as the deformation becomes significant -- 
the loss of the closed-shell structures results in
a more Pauli-like behavior of the total spin\cite{Eto}. 

Fig. 3 shows the up-spin densities in deformed 
potentials for $N=31, N^{\uparrow}=16$.
We find that true CDW ground states are induced by the increasing
elliptical 
deformation of the external potential.
The ground-state spin is  
$S=\displaystyle{1/2}$ in all three cases.
The charge-density wave has period $\ \sim \cos^2(5 \theta)$ and 
results from the mixing of the degenerate $L_z=\pm 5$ states
by the elliptical external potential.

In conclusion, we have studied the electronic states of quantum dots
with up to
58 electrons for parabolic circular and elliptical external potentials,
using spin-density-functional theory and exact diagonalization.
For a circular potential, we observe a shell structure for the 
filling of the dot with electrons. Hund's rule determines the spin 
configuration of the ground state up to 22 electrons.
For specific numbers of electrons, CDW states appear on small 
elliptical deformation of the
external 
potential, while previously identified SDW states \cite{Koskinen} are
found to 
be artifacts of broken spin symmetry in density-functional theory.
For elliptical potentials, the 
shell structures are lost with increasing deformation,
and the spin configurations change from Hund's rule to a more Pauli-like
behavior. 

We acknowledge O.Agam, I.L.Aleiner, B.L.Altshuler, D.J.Chadi, Y.Meir,
W.Kohn, and M.Stopa for 
comments and suggestions.

\newpage

$~\dagger$ Permanent address: Fundamental Research Laboratories, NEC
Corporation, 34 Miyukigaoka, Tsukuba, 305-8501, Japan\\

\newpage
\noindent
{\bf Figure Captions}
\begin{itemize}

\item[Figure 1:] Addition energy $\Delta(N)$ as a function of electron
number $N$ in the dot for
(a) a confining parabolic potential
$\displaystyle{m^*\omega_0^2r^2/2}$ and (b,c) elliptical
confining potentials
$\displaystyle{V({\bf r})=m^*(\omega^2_xx^2+\omega^2_yy^2)/2}$.
The dotted lines indicate the addition energy
according to a classical electrostatic analysis\cite{Shikin}.
The parameters are (a) $\hbar\omega_0=3.0$ meV,
(b) $\displaystyle{\omega^2_y/\omega^2_x=11/13}$
and (c) 
 $\displaystyle{\omega^2_y/\omega^2_x=5/7}$.
(b) and (c) are shifted by 1.0 meV and 2.0 meV, respectively.\\
Upper inset - total energy in meV as a function of total spin for
electron
numbers $N=16,24,34$, and $46$ in (a). The origin of energy for each $N$
is arbitrary.\\
Lower inset - total energy for $N=4$ electrons obtained by exact
diagonalization within a restricted Hilbert space, $|0,\pm 4\rangle$
and $|1,\pm 2\rangle$, as a function of single-particle level splitting
$\Delta$.
The energies  
$E(S=0,L_z=0)$ and $E(S=1,L_z=\pm 2)$
are plotted relative to $E(S=2,L_z=0)$.\\

\item[Figure 2:] Ground-state spin as a function of electron number
$N$ for (a) the parabolic confining potential and (b,c)
the elliptical confining potentials of Fig. 1
The dotted line in (a) indicates the spin
configuration when Hund's rule is satisfied. 

\item[Figure 3:] Charge-density distributions at $N=31$.
Left column - from top to bottom, spin-up ($N^{\uparrow}=16$),
spin-down ($N^{\downarrow}=15$), and total charge-density distribution.
Right column - spin-up charge-density distributions ($N^{\uparrow}=16$)
for elliptical potentials, 
$\displaystyle{\omega_y^2/\omega_x^2= 11/13}$ (top), 
$\displaystyle{\omega_y^2/\omega_x^2= 4/5}$ (middle), 
and $\displaystyle{\omega_y^2/\omega_x^2= 5/7}$ (bottom), 
respectively.

\ \\
\ \\
\ \\

\end{itemize}

\newpage

{\LARGE\bf Table 1}

\vspace{2.0cm}

\begin{tabular}{c|ccc}\hline\hline
{\rm Number of Electrons (N) }& {\rm spin up $({\rm N}^{\uparrow})$}&
                 {\rm spin down $({\rm N}^{\downarrow})$}& {\rm total
$Lz$} \\ \hline
 3   &   {\bf 2}   &         1   & 1  \\ 
 5   &        3    &   {\bf  2}  & 1  \\
 7   &   {\bf 4}   &         3   & 2  \\
10   &        6    &   {\bf  4}  & 2  \\
13   &   {\bf 7}   &         6   & 3  \\
17   &       10    &   {\bf  7}  & 3  \\
21   &  {\bf 11}   &        10   & 4  \\ 
23   &       12    &   {\bf 11}  & 4  \\ 
31   &  {\bf 16}   &        15   & 5  \\
33   &       17    &   {\bf 16}  & 5  \\ 
43   &  {\bf 22}   &        21   & 6  \\
45   &       23    &   {\bf 22}  & 6  \\
57   &  {\bf 29}   &        28   & 7  \\ \hline\hline

\end{tabular}

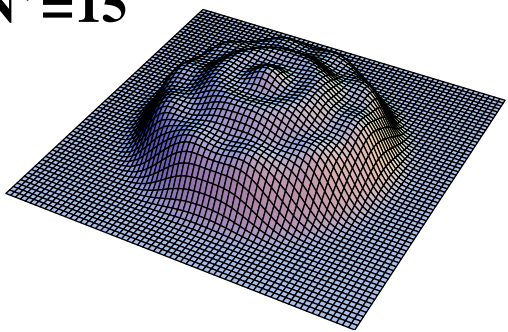
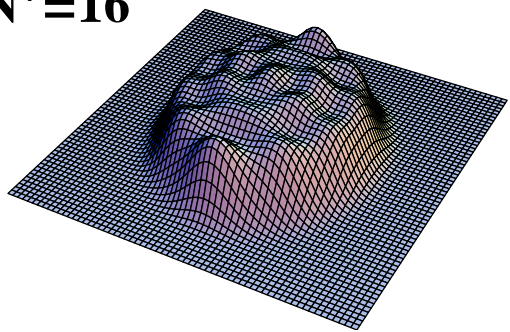
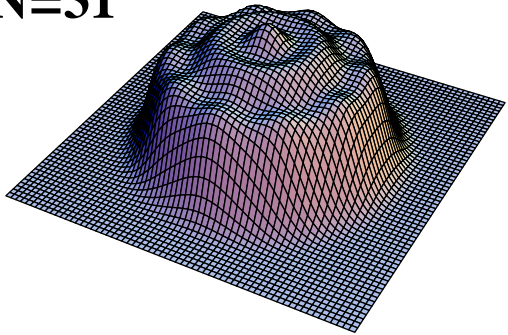
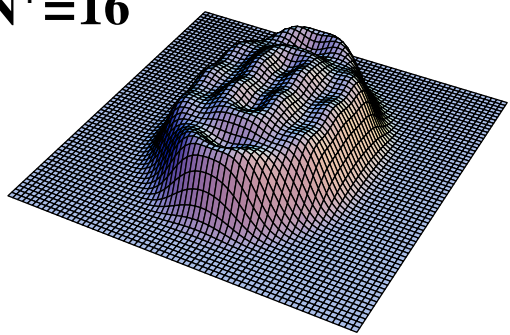
\end{document}